\documentclass{article}

\usepackage{arxiv}

\usepackage[utf8]{inputenc} 
\usepackage[T1]{fontenc}    
\usepackage{hyperref}       
\usepackage{url}            
\usepackage{booktabs}       
\usepackage{amsfonts}       
\usepackage{nicefrac}       
\usepackage{microtype}      
\usepackage{lipsum}
\usepackage{amsmath,amsfonts,amssymb}
\usepackage{caption} 
\usepackage{subcaption} 
\usepackage{graphicx}
\usepackage{pgfplots}
\usepackage[all]{nowidow}
\usepackage[utf8]{inputenc}
\usepackage{tikz}
\usetikzlibrary{er,positioning,bayesnet}
\usepackage{multicol}
\usepackage{algpseudocode,algorithm,algorithmicx}
\usepackage[inline]{enumitem}
\usepackage{adjustbox}

\definecolor{blue}{HTML}{1F77B4}
\definecolor{orange}{HTML}{FF7F0E}
\definecolor{green}{HTML}{2CA02C}

\pgfplotsset{compat=1.14}

\title{Selectivity correction with online machine learning}

\author{
  Max Halford \\
  IRIT Laboratory \\
  IMT Laboratory \\
  University of Toulouse \\
  \texttt{max.halford@irit.fr} \\
   \And
 Philippe Saint-Pierre \\
  IMT Laboratory\\
  University of Toulouse \\
  \texttt{philippe.saint-pierre@math.univ-toulouse.fr} \\
   \And
 Frank Morvan \\
  IRIT Laboratory \\
  University of Toulouse \\
  \texttt{frank.morvan@irit.fr} \\
}

\begin{document}
\maketitle

\begin{abstract}
Computer systems are full of heuristic rules which drive the decisions they make. These rules of thumb are designed to work well on average, but ignore specific information about the available context, and are thus sub-optimal. The emerging field of machine learning for systems attempts to learn decision rules with machine learning algorithms. In the database community, many recent proposals have been made to improve selectivity estimation with batch machine learning methods. Such methods are all batch methods which require retraining and cannot handle concept drift, such as workload changes and schema modifications. We present online machine learning as an alternative approach. Online models learn on the fly and do not require storing data, they are more lightweight than batch models, and finally may adapt to concept drift. As an experiment, we teach models to improve the selectivity estimates made by PostgreSQL's cost model. Our experiments make the case that simple online models are able to compete with a recently proposed deep learning method. 
\end{abstract}


\section{Introduction}

Heuristics are aplenty throughout computer systems. Applications possess many parameters that can be tuned in order to improve a measure of performance. For instance, page sizes in relational databases \cite{weisberg2009using}, cache sizes in storage systems \cite{oh2012caching}, and backoff amounts for re-transmits in networking \cite{kwak2005performance}, are all parameters that are chosen somewhat heuristically. A significant amount of work has explored the idea of replacing these heuristics with rules found by a machine learning algorithm. For instance, \cite{mirhoseini2017device} proposed a reinforcement learning method for optimising device placement in a heterogeneous distributed environment which outperformed human experts. Meanwhile the authors of \cite{kraska2018case} proposed to replace index structures such as B-trees with what they call \emph{learned indexes}. What ties all these proposals together is that they each use machine learning to replace a heuristic method. Whereas heuristics are chosen because they work well on average, a machine learning algorithm can learn to use the available context of every specific case, and may thus outperform a heuristic. Moreover, a machine learning algorithm can improve through time and exploit usage patterns. Another consequence is that the resulting systems may be easier to maintain because they require less code written by humans. This has become a prevalent research topic that has been coined \emph{machine learning for systems} \cite{dean2017machine}.

This trend has began to seep into the database query optimisation community. In a relational database, a query optimiser is tasked with choosing an execution plan which answers a query in the least amount of time possible. The quality of a query optimiser is mostly based on its ability to predict the cost of each candidate execution plan. The query optimiser relies on a cost model in order to do so. Database cost models have been using heuristics that have evolved ever so slightly since the seminal work of \cite{selinger1979access}. Cost estimation -- of which selectivity estimation is a sub-task -- is therefore prone to large mistakes that deteriorate the quality of the query execution plans \cite{leis2015good}. In recent years, there has been a regain in interest from the database community to explore the use of machine learning to improve query optimisation. While some have taken the extreme approach of learning a query optimiser from scratch \cite{ortiz2018learning,marcus2019neo}, most have focused on improving the accuracy of the cost model. They all follow the same approach, which is to predict the cost of a query execution plan (QEP) by training models on a history past QEPs. Some have focused on solely predicting a query's selectivity \cite{wu2018towards,kipf2019estimating,woltmann2019cardinality}, while others directly model its cost \cite{akdere2012learning,marcus2019plan}.

We believe that replacing heuristics with machine learning -- and in particular deep learning -- is a promising solution towards improving cost models. However, we posit that the methods referenced hereabove are flawed in certain ways. First of all, every proposed method uses a batch perspective, whereby a model is trained on a large history of queries. A batch model is static, and cannot learn from new queries without being retrained from scratch. The second issue is that the proposed methods are not able to handle database schema changes as well as modifications in access patterns. On the contrary, batch models assume that the queries seen during the offline training phase are representative of the queries that will be seen during the online prediction phase. However, in practice, attributes and relations can be added, updated, and dropped. Attribute distributions and correlations might also change through time. Additionally, query workloads are susceptible to evolve. We put all of these changes through time under the umbrella term \emph{concept drift} \cite{gama2014survey}. A batch model is able to handle concept drift by regularly being retrained. Alas, it will essentially be playing a cat-and-mouse game, as it is not able to constantly stay up to date. Finally, the recent trend towards deep learning isn't realistic for efficiency reasons; simpler methods are therefore still worth considering.

We propose an alternative approach based on \emph{online machine learning} \cite{10.1145/3373464.3373470}. Under this regime, a machine learning model is able to update itself with every incoming observation. An online model has the \emph{anytime property}, meaning that it can produce predictions at any moment in its lifetime. Moreover, online models can adapt to concept drift. To showcase the benefits of online machine learning, our work in this paper focuses on the task of selectivity estimation. Specifically, our goal is to correct an existing selectivity estimation module by learning to predict the mistakes it will make. Our method can therefore be plugged into any selectivity estimation module. Our contributions are as follows: 1) we formulate selectivity estimation as an online machine learning problem and enumerate the resulting advantages over a batch learning approach, 2) we demonstrate how a small set of features as well as a simple model are as accurate and more efficient than MSCN \cite{kipf2019estimating}, which is a batch model based on neural networks, and 3) we show that an online machine learning model is able to cope with concept drift and may thus outperform a batch model as time goes on.

\section{Background and related work}

\subsection{Selectivity estimation}

Whenever a user issues a query to a database, the query optimiser is tasked with finding the optimal query execution plan (QEP). The total query time, as perceived by the user, is the sum of the query optimisation time and the query execution time. The query optimiser thus has to compromise between finding an efficient QEP and not spending too much time doing so. To choose a QEP, the query optimiser enumerates a set of candidate plans, estimates the cost of each plan, and picks the one with the lowest cost. A QEP is a tree of physical operators (e.g.\ joins, scans, filters). Therefore, in a single processor environment with no communication cost, the cost of a QEP is the sum of the cost of each of its operators. The cost of a physical operator is largely determined by the amount of tuples that flow from the operators that precede it. This amount is called the \emph{selectivity}. To this end, the cost model has access to \emph{metadata} that summarises the distribution of the data. The issue is that storing a distribution of multiple attributes takes an amount of space that grows exponentially with the number of attributes. In practice, the \emph{attribute value independence} (AVI) assumption is made, which enables the cost model to assume that attributes are independent with each other. Therefore, the cost model only has access to individual attribute distributions. The cost model makes other simplifying assumptions, such as the \emph{join uniformity assumption}, which states that attribute value distributions are preserved following a relational join.

Selectivity estimation is considered to be the most important part of the cost model \cite{gu2012testing,leis2015good}. Alas, the simplifying assumptions that are usually made lead to estimates that are off by several orders of magnitude. Indeed, estimation errors made early on in the QEP can grow exponentially and have devastating consequences further up the plan \cite{ioannidis1991propagation}. In turn, said errors will deteriorate the accuracy of the cost model \cite{lohman2014query}, which in turn worsens the query optimiser's performance. However, these assumptions allow the selectivity estimation process to run in a very short amount of time, which is of paramount importance. Therefore, a lot of research has delved into relaxing said assumptions, at as little a cost as possible. A first area of focus has been on improving the accuracy of the one-dimensional attribute value distributions, i.e.\ \emph{attribute level synopses}. In database cost models, histograms are ubiquitous \cite{poosala1996improved,ioannidis2003history}. Estimating the distribution of a single attribute is essentially a solved problem. However, attribute level synopses do not help whatsoever in capturing dependencies between attributes, which are the Achilles Heel of cost models \cite{lohman2014query}. Consequently, \emph{table-level synopses} have been proposed. This includes extensions of histograms to multiple dimensions \cite{muralikrishna1988equi,bruno2001stholes}, as well as Bayesian networks \cite{tzoumas2011lightweight,halford2019approach}. Additionally, significant efforts have been made into developing sampling methods, both for single relations \cite{riondato2011vc} and multiple relations \cite{vengerov2015join}. However, apart from in-memory databases, sampling imposes a high computational cost and in general suffers from the empty-join problem \cite{chaudhuri1999random}.

Selectivity estimation is a difficult problem. The simplifying assumptions made by the cost model are difficult to relax without storing an inordinate amount of metadata. Cost models sometimes use heuristic formulas in order to soften these assumptions without introducing any additional complexity. These heuristics are intended to work in general cases, but have no guarantee of being optimal for a given database and a particular workload. As an alternative, a growing body of research has explored the idea of \emph{learning} to predict the selectivity estimation based on past QEP execution feedback and contextual information.

\subsection{Learning to estimate selectivities} \label{learning-to-estimate-selectivities}

Once a QEP has been picked by the query optimiser, it is executed and the results are delivered to the user. When the execution is finished, the true selectivity at each stage of the QEP is made available. Therefore, it is possible to measure the error of each selectivity estimate every time a query is executed. Early on, the authors of \cite{aboulnaga1999self} proposed to exploit the feedback to tune the parameters of both univariate and multivariate \emph{query-driven histograms}. Meanwhile, \cite{stillger2001leo} were the first to propose an adaptive cost model which can learn from feedback for any kind of QEP. Their system, named LEO -- for ``LEarning Optimizer'' -- is based on the simple idea of memorising the exact selectivities of frequently occurring operators. Their approach can thus be resumed with the saying ``once bitten, twice shy''. Naturally, a memorising approach such as LEO does not help whatsoever when having to determine the selectivity of unseen queries.

The issue with LEO is that it uses the raw representation of a QEP. If a new QEP differs in any way from each memorised QEP, then LEO is essentially blind. A more sophisticated approach is to project QEPs into a new domain wherein similar QEPs have similar traits. In practical terms, the idea is to represent a QEP by a set of \emph{features}. The goal is to define features that are correlated with the query selectivities. A machine learning algorithm can then be taught to map the features to the selectivities. In the case of selectivity estimation, \emph{linear models} have been proposed \cite{wu2018towards}, as well as support vector machines (SVM) \cite{akdere2012learning}, and neural networks \cite{liu2015cardinality}. Independently of the model choice, the challenge is to find meaningful features which help in accurately predicting the selectivity of a QEP. Historically, designing good features for machine learning purposes has been done manually by domain experts.

In recent times, there has been a surge of proposals that aim to automatically learn these feature via a neural network. The authors of \cite{kipf2019estimating} introduced \emph{MSCN}, which is a \emph{convolutional neural network} (CNN) that takes as input a \emph{one-hot encoded} version of a QEP. That is, they convert an input query to a set of three vector of 1s and 0s. Each vector indicates the presence (and absence) of each existing attribute, join key, and relational operator. They then stack a few linear layers with rectified linear units and train the neural network to map the binary vectors to selectivities. In their experiments, they manage to outperform the accuracy of PostgreSQL's cost model by a factor of 3. In order to lower the high computational cost required by their method, \cite{woltmann2019cardinality} explored the idea of breaking down their single neural network into many smaller ones. Each sub-network is in charge of capturing the dependencies at specific parts of the database schema. Indeed, they surmise that attribute dependencies between distant relations are less important than between adjacent relations. Other methods based on deep learning have also been proposed. For instance, \cite{dutt2019selectivity} have proposed a specific neural network designed to handle range predicates over numerical variables.

Overall, learning to estimate selectivities seems to be a practical method with promising results. Machine learning methods seem to regularly outperform established methods, at least according to recent research papers. This is no surprise, as established methods are traditionally simplistic. The reasons why they are used in the first place is because they are very efficient and do not require an expensive learning phase. Meanwhile, machine learning models imply heavier algorithms that require a learning phase where all the training data has to be available. Moreover, batch models cannot be dynamically updated. For instance, the one-hot encoding schemes from \cite{kipf2019estimating} and \cite{woltmann2019cardinality} cannot cope when an attribute is added to a relation. Finally, batch models cannot deal with concept drift. For instance, they do not account for distribution modifications tuples are inserted or removed from the database, and cannot adapt when the query workload evolves. Therefore, there is some room for improvement in being able to handle the never-ending stream of incoming queries. Because of the streaming nature of the problem, we believe that using an online model which can handle such a stream is a more adequate solution. Models that can do so are part of a sub-field of machine learning named \emph{online machine learning}. We will now give an overview of the latter, before framing it in the case of selectivity estimation.

\section{Methodology}

\subsection{The benefits of online machine learning}

Machine learning involves teaching a computer program to learn to perform a task by showing it examples. In a regression task, such as selectivity estimation, the computer program has to learn a function $f : {\rm I\!R}^p \rightarrow {\rm I\!R}$. That is, for a given set of $p$ features $x$, the program is capable to output a numeric prediction $\hat{y} = f(x)$. The performance of $f$ can be measured by comparing the prediction $\hat{y}$ with the true outcome $y$, which, in a live environment, is unknown at the time of making the prediction. Machine learning models traditionally operate in a batch setting, whereby a bunch of $n$ feature sets $X = \{x_1, \dots, x_n\}$ and $n$ known outcomes $Y = \{y_1, \dots, y_n\}$ are available. During the training phase, the program learns a function $f$ from the so-called training set $(X, Y)$. During the prediction phase, the function $f$ is applied to new sets of features. It is implicitly assumed that the training set and the test set possess the same statistical distributions, which isn't necessarily the case in practice. The main issue with a batch learning model is that it has to be retrained from scratch in order to learn from new data. Indeed, the predictive function $f$ produced by a batch learner is set in stone and cannot be updated incrementally with a new observation $(x, y)$. However, in a live environment, a never-ending stream of training data is constantly being produced. In such a scenario, $f$ is immediately sub-optimal because it isn't exploiting the new observations. This is even more so the case when concept drift occurs, which implies that the non-stationary condition of the data distribution is being violated. In this case, the model's performance is susceptible to plummet. In practice, batch learning models are periodically retrained. The choice of the period between training sessions is a compromise between the cost of the training and the regret in terms of missed opportunity. Each of the methods mentioned in subsection \ref{learning-to-estimate-selectivities} follow this paradigm.

In contrast, an online machine learning model can be updated every time a new observation arrives. It also has the \emph{anytime} property, meaning that it can make predictions at any moment in its lifetime. An observation can be discarded once it has been used to update $f$. This allows not having to store a training set, which simplifies things from a data engineering point of view. Moreover, an online learning model is \emph{dynamic}, in that the number of features can be altered on-the-fly. Indeed, an online learning model is resilient and will still work if features are added or are not available. Thirdly, an online model is capable of being \emph{adaptive}, meaning that it can adjust itself when concept drift occurs. Last but not least, the performance of the model can be measured in real-time. Indeed, a performance metric can be maintained by comparing the model's prediction with the ground truth once it is available, \emph{before} updating the model. This is called \emph{progressive validation} \cite{blum1999beating} and allows all the data to be used both as a training set as well as an evaluation set. Many approaches have been proposed for online machine learning, including partition trees and linear models. We refer readers to \cite{10.1145/3373464.3373470} and references therein for a comprehensive review. Online machine learning is a strong fit for selectivity estimation because of the streaming aspect of the problem. We will now formalise our approach and describe our methodology. We will then explain how we convert QEPs to features and provide an overview of the online machine learning models we considered.

\subsection{Learning correction factors} \label{correction-factor}

Cost models used in modern databases are known to be highly error-prone. Specifically, they are susceptible to underestimate selectivities by a significant amount \cite{lohman2014query}. Practitioners resort to heuristic rules in order to alleviate this issue without introducing too much computational burden. For instance, take the case where multiple \texttt{WHERE} predicates occur on a single relation. The simple way is to estimate the selectivity of every predicate separately and multiply them with each other afterwards. This approach is naive, as it is based on the attribute value independence assumption. In order to soften the latter, one of SQL Server's heuristics is instead to take the minimum of the selectivities. By doing so, SQL Server assumes a worst-case scenario whereby the predicates are perfectly correlated. This heuristic is justified because of the fact that, supposedly, it works well for the workloads of SQL Server's user base.

We can introduce a simple heuristic to fix the underestimation of a cost model. We denote by $y$ the true selectivity, which is available once the query has been successfully executed. Meanwhile, $\hat{y}$ denotes the estimated selectivity. For a workload of $n$ queries, the average underestimation amount, which we denote by $c$, is equal to $\frac{1}{n} \sum_{i=1}^n \frac{y_i}{\hat{y}_i}$. We can reduce the amount of underestimation for future queries by multiplying their associated selectivities by $c$. The latter therefore acts as a correction factor which can be updated every a query is processed. This heuristic rule can be refined by increasing its granularity. For instance, we can segment queries according to the number of joins they require and keep track of the multiplicative error made within each segment. Therefore, instead of having a single correction factor $c$, we would have one correction factor $c_j$ for each number of joins $j$. A natural extension is to learn to predict the correction factor based on the characteristics of each particular query execution plan. This is a straightforward case of supervised regression learning which can be done in an online manner.

\subsection{Extracting useful features from a plan}

A query execution plan as such cannot be processed by a machine learning model. Indeed, most models are designed to process numeric features. As per usual in machine learning, the goal is to find features that are correlated with the output and uncorrelated with each other. The key advantage of deep learning methods is that they do not have to manually define features. Instead, they learn features by starting off from the one-hot encoded representation of the QEPs. As we explain in the next subsection, this is akin to using a factorisation machine \cite{rendle2010factorization}. Our feature extraction process is based on the observation that there are two types of QEPs we may encounter. On the one hand, a QEP might already have been seen in the past. Indeed, it is quite common for queries to be repeated by users. Moreover, parts of each QEP might have been encountered in the context of other queries. This redundancy is a low-hanging fruit that is the basis of DB2's LEO optimiser \cite{stillger2001leo}. On the other hand, some QEPs might never have been seen before. In this case, memory-based approaches such as LEO are useless. Because of this duality, we defined two sets of features.

The first set of features is meant to address QEPs which have never been seen before. In this case we record general information: 1) the number of joins, 2) the number of involved relations 3) the number of \texttt{WHERE} statements, 4) the maximum number of \texttt{WHERE} clauses on a single relation. The second set of features is aimed at exploiting redundant QEPs. We calculate average selectivity estimation errors by grouping over various information from each QEP. In machine learning, this is referred to as \emph{target encoding} \cite{micci2001preprocessing}. The idea is to extract information such as the relations, joins, attributes, attribute values, and replace them by the average of the selectivity estimation errors encountered in the past. For example, if a QEP contains a filter condition on a \texttt{country} attribute, then we would look at the average selectivity estimation error of queries which involved it. We can also capture dependencies by target encoding on combinations, such as on pairs of attributes. This can be done in a streaming fashion by noticing that an average can be updated online: $\bar{x}_{i+1} = \bar{x}_i + \frac{x - \bar{x}_i}{i+1}$ where $\bar{x}_i$ is the current average and $x$ is the new value (in our case, a selectivity). The main issue of target encoding is that its outputs are somewhat unreliable early on. Indeed, if a particular attribute has only been used in one particular QEP before, then the error that was made for that QEP will be the only one included in the average. To alleviate this issue, one may use a Bayesian average, which requires choosing a prior value and a weighting term in order to compute a weighted average between the observed average and the prior value.

\subsection{Choice of online learning models}

The most common approach to online machine learning is to use differentiable models that minimise an empirical loss function. For instance, a common objective in regression is to find the parameters $\theta$ of a model $m$ which minimise the squared loss between the training set outputs $y$ and the predicted outputs $\hat{y} = m(x)$ made over the training set inputs $x$:

\begin{equation}
    \underset{\theta}{min} \sum_{i=1}^{n} (y_i - m(x_i)) ^ 2
\end{equation}

If $m$ can be differentiated with respect to its inputs, then for each pair $(x_i, y_i)$ a gradient $g_i$ can be obtained. The gradient can then be used to update $\theta$. This is called \emph{stochastic gradient descent} \cite{bottou2012stochastic} as only one pair is being processed at a time. In the case of a linear regression, $\theta$ will correspond to the weights assigned to each input $x_i$. Meanwhile in the case of a neural network, $\theta$ corresponds to the weights and biases of each neuron at each layer.

Stochastic gradient descent comes in different flavors. The simplest variant is to multiply the gradient by a learning rate $\eta$ and subtract the result from $\theta$. The gradient provides the update direction as well as the its magnitude, whilst the learning rate acts as a regulariser for the magnitude. In order to obtain convergence, a common trick is to define a schedule which reduces the learning rate as time goes by \cite{bottou2012stochastic}. Although learning rate schedules are commonly used in practice, they are ill-suited for online learning. Indeed, if we expect to observe drift, then lowering the learning rate means that the model will become less capable of adapting. In our experiments we thus decided to use a constant learning rate. We refer to \cite{ruder2016overview} for an overview of more sophisticated flavors. Note that we considered Online Newton Step (ONS) \cite{hazan2007logarithmic}, which is a second-order optimisation method that approximates the Hessian of the loss function, but discarded it because of speed concerns.

Many models can be trained with stochastic gradient descent. The simplest one is linear regression, which is just a dot product between a set of $p$ weights $w_i$ and a set of $p$ input features $x_i$: $\hat{y}_i = \sum_{j=1}^p w_{ij} x_{ij}$. Although linear regression is a simplistic method, it has the advantage of being fast and interpretable. Neural networks (including deep learning) can also be trained via stochastic gradient descent, and in fact almost always are. In our experiments, we only considered the use of standard feed-forward networks. These can be seen as a succession of dot-products $f$ interleaved by non-linear activation functions $a$ (e.g., sigmoid, tanh, ReLU): $\hat{y} = a(f(a(f \dots a(f(x)))))$. In a neural network, the gradient can be obtained by working backwards from the loss function and accumulating the individual derivatives at each step along the way. This process is the well-known \emph{back-propagation} algorithm. Neural networks may be preferable to linear regression because they take into account interactions between features.

Another way to take into account feature interactions is to compute polynomial combinations of said features and feed these to a linear regression. For a given parameter $k > 1$, one can thus obtain $\sum_{i=2}^k p^k$ additional features, where $p$ is the cardinality of the features. This provides a simple way to give non-linear capacity to a linear model. The downside is that the number of additional features can grow extremely large, which in practice is a strong deterrent. Moreover, some feature interactions might occur very rarely, which may lead to uncertain parameter estimation and over-fitting. Factorisation machines (FM) \cite{rendle2010factorization} have been proposed as to circumvent these issues. Instead of explicitly estimating a parameter for each combination of features, a FM stores \emph{latent parameters} for each feature. To quote the original FM paper, ``\emph{instead of using an own model parameter $w_{i,j} \in \mathcal{R}$ for each interaction, the FM models the interaction by factorising it}''. Each feature $x_j$ is thus associated with a latent vector $v_j$ of length $k$. A latent vector therefore contains $k$ values which describe the associated feature. Intuitively, features used in the same context, such as \texttt{nationality = Japanese} and \texttt{country = Japan} will have similar latent vectors. The interaction between two features is then computed as the dot product between their associated latent vector:

\begin{equation}
    \hat{y_i} = w_0 + \sum_{j=1}^p w_j x_j + \sum_{j=1}^p \sum_{k=i+1}^p \langle v_j,v_k \rangle x_j x_k 
\end{equation}

We can now store $k \times p$ weights instead of $p^2$. Even if two features have never observed together in a single observation, their interaction parameter can be estimated from their respective latent vectors. FMs have been used with great success for high-dimensional problems such as click-through rate prediction \cite{pan2018field}. They suit the selectivity estimation problem because of the high number of attributes, values, relations, and joins present in a database schema. In particular, FMs are an elegant way to model attribute dependencies.

Decision trees are another class of models which are able to capture dependencies. They are based on a simple principle, which is to partition the space of observations into boxes, whereby the observations contained inside each box are homogeneous. The idea is to recursively partition the data by finding split rules which minimise a heterogeneity criterion, such as entropy for classification or mean squared error for regression. Decision trees are usually trained in a batch manner \cite{breiman2017classification}. However, there have been a few proposals to train them online. The most established method is called the Hoeffding tree \cite{domingos2000mining}. Because all the data is not available at once, the idea is to maintain summary statistics $P(x_i \mid y_{ik})$ where $x_i$ are features and $y_{ik}$ is a label. The summary statistics are stored inside each leaf of the tree and are updated each time an observation is sorted into a leaf. The tree starts off as a single leaf. Every so often, the summary statistics are used to evaluate possible split rules. A leaf is split into two leaves, and thus becomes a branch, once the gain from the heterogeneity criterion surpasses a certain threshold, which is called the Hoeffding bound. A Hoeffding tree is guaranteed to find the same structure as a batch decision tree as long the underlying data distribution is stationary.

Finally, we made a minor contribution by developing our own online machine learning model. We adapted Bayesian linear regression to make it robust to concept drift. In a nutshell, Bayesian linear models attach uncertainty to each weight. As more data arrives, they provide a mathematical framework for updating each weight whilst accounting for the current uncertainty and the information brought by the new data. Although sound by design, such models assume a stationary environment, and do not cope well with concept drift. To circumvent this, we have introduced a variant to Bayesian linear regression which is based on exponential moving averages. As far as we are aware, it has not been proposed in existing publications. However, the focus of this paper is to motivate the use of online machine learning models, and not necessarily to propose a state-of-the-art solution. Therefore, we leave the details in appendix \ref{robust-bln}, and reserve ourselves the right to explore the matter in further depth in a subsequent publication.

\section{Evaluation}

We evaluated our approach using the IMDb dataset from the JOB benchmark \cite{leis2015good}. The IMDb dataset contains real-word data pertaining to the movie industry and contains many correlations (for instance French actors usually play in French movies). It contains 21 relations and weighs 3.6 GB. We simulated query workloads by sampling from the queries that accompany each dataset. In the case of IMDb there are 113 available queries. For each sampled query, we asked PostgreSQL to generate an execution plan and sampled sub-plans from each of these execution plans. For each sub-plan, we tasked each machine learning model we benchmarked to predict the correction factor we introduced in subsection \ref{correction-factor}. We then multiplied each predicted correction factor with the selectivity estimate made by PostgreSQL's cost model. Finally, we calculated the $q$-error \cite{moerkotte2009preventing}, which measures the multiplicative difference between predictions and ground truths, and has become the \textit{de facto} standard for evaluating the quality of selectivity estimation methods.

We benchmarked the following online models: \begin{enumerate*} \item Linear regression trained with SGD and a learning rate of 0.1 \item Hoeffding tree with a patience of 200 and a maximum depth of 5 \item Feed-forward neural network with 2 hidden layers of 30 neurons trained with Adam and a learning rate of 0.01. \item FM with 10 components trained with SGD and a learning rate of 0.1. We provided it with one-hot encoded versions of the used attributes, attribute values, relations, and joins. \item Bayesian linear regression with $\gamma = 0.7$ \end{enumerate*}. As a comparison, we included the selectivity estimates made by PostgreSQL's cost model. We also benchmarked the following batch learning methods: \begin{enumerate*} \item Standard linear regression fitted with maximum likelihood estimation (MLE). \item LightGBM \cite{ke2017lightgbm} with default parameters, which is probably the best off-the-shelf batch learning algorithm, and therefore provides an interesting reference. \item MSCN from \cite{kipf2019estimating}, which we trained for 50 epochs. \end{enumerate*}

\begin{figure}[H]
    \centering
    \scalebox{0.53}{\includegraphics{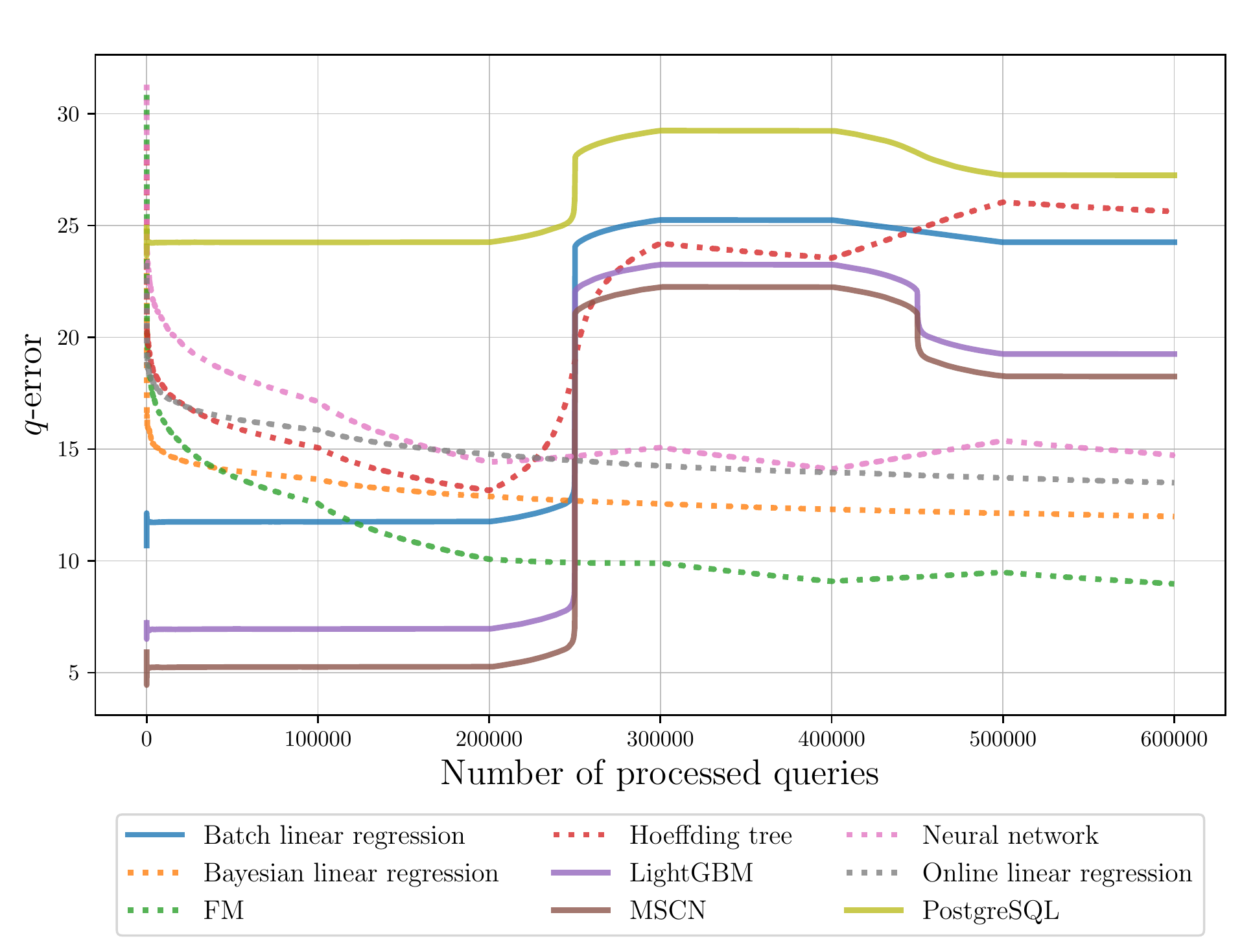}}
    \caption{$q$-errors for each method with hard concept drift}
    \label{fig:q-errors-drift-hard}
\end{figure}

We trained each batch method on a 100,000 execution plans prior to conducting the benchmark. Meanwhile, the online methods do not require this warm-up phase because they are trained online. The evaluation phase samples 600k query execution plans and obtains a prediction from each method. In order to simulate concept drift, we clustered the queries into three buckets. Queries within each bucket operate on similar relations and thus possess some similarities. Intuitively, a model that is trained on one bucket will have difficulties estimating selectivities for queries from other buckets. We swapped buckets after 200k and 400k queries, therefore establishing two hard concept drifts. Figure \ref{fig:q-errors-drift-hard} shows a running average of the $q$-errors for each method along time. Online models are represented with dotted lines, whilst solid lines are used for batch models. As can be expected, the batch methods initially outperform the online methods because they have has a warm-up phase. However, in the online learning methods eventually outperform their batch counterparts, in particular because they are able to adapt to the hard concept drift, except in the case of the Hoeffding tree. The best performing method is FM. Naturally, batch learning models can be retrained in order to cope with concept drift, but that would require storing observations and defining a training schedule, which isn't particularly trivial to put in place.

\begin{figure}[H]
    \centering
    \scalebox{0.53}{\includegraphics{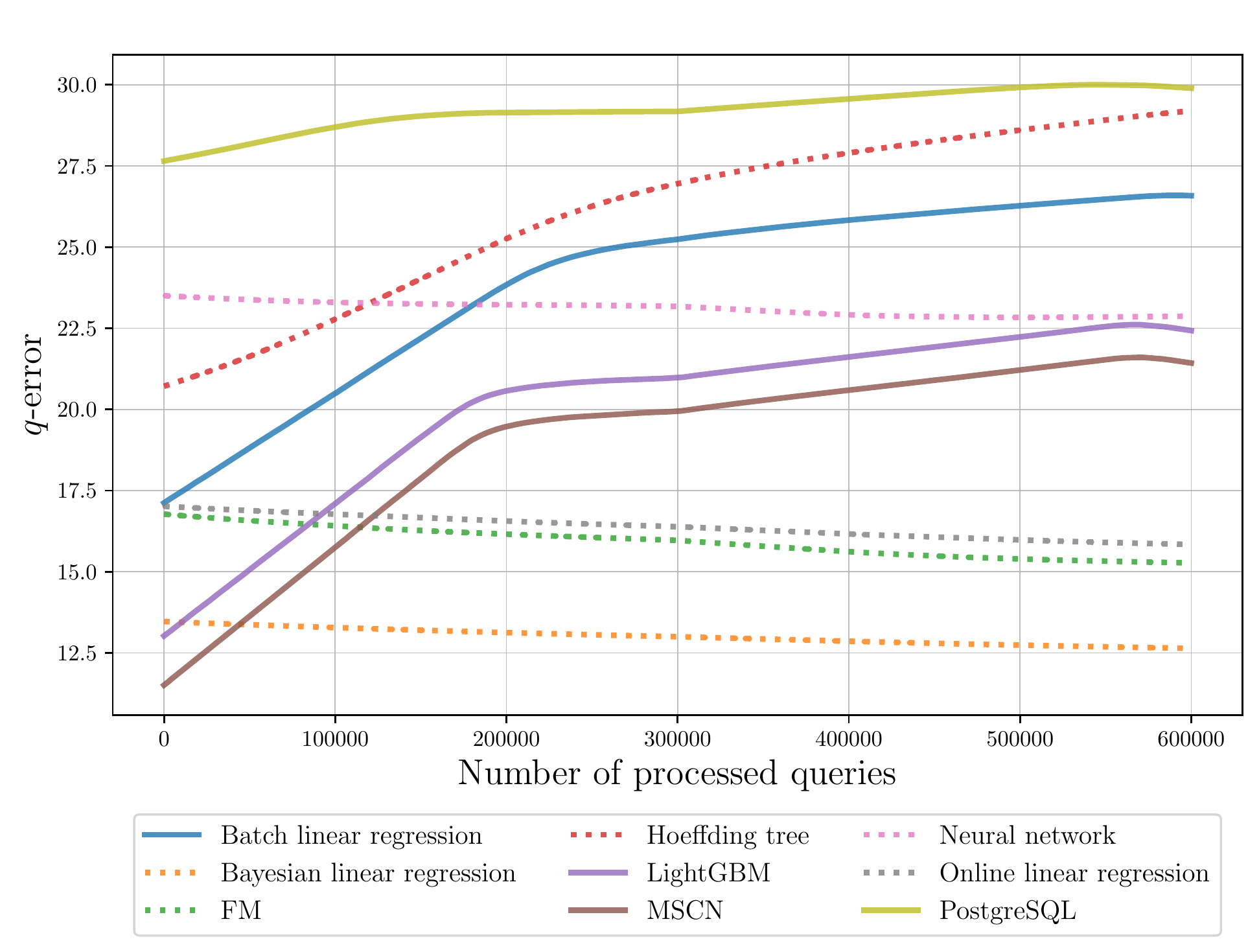}}
    \caption{$q$-errors for each method with soft concept drift}
    \label{fig:q-errors-drift-slow}
\end{figure}

We also experimented with slow concept drift. Instead of switching buckets at predetermined moments, we decide which bucket to sample from in a probabilistic manner. At every moment $t$, the probability of sampling from bucket $b$ is:

\begin{equation}
    P(b, t) = \frac{exp(-\frac{(t - t_b)^2}{d})}{\sum_{b_i} exp(-\frac{(t - t_{b_i})^2}{d})}
\end{equation}

where $t_{b}$ are predetermined values that we spread out in a uniform manner (i.e. 150k, 300k, 450k respectively for each bucket). Meanwhile, $d$ determines the abruptness of the drift (i.e. a large $d$ corresponds to a harder drift). For our experiments we arbitrarily chose $d = 3$. The results of this simulation are provided in figure \ref{fig:q-errors-drift-slow}. As can be seen, the performance changes are smoother than in the hard drift case, which is possibly more representative of what may occur in the real world. The performances of the batch models aren't good because they have been trained on data from the first bucket. The online models are all able to cope, except for the Hoeffding tree. However, the ranking of the models isn't the same. Indeed, the best performing model is the Bayesian linear regression, whereas the FM model comes second. This could be due to the fact that it contains more weights that have to be modified, which is also the case for the neural network.

\section{Conclusion}

Computer systems, including database systems, have to make many choices that affect their performance. These choices often involve heuristic decision rules. For instance, database cost models use dampening tricks to soften the attribute value independence assumption. Said rules are designed to work well on average, but do not exploit the available context. Providing an automatic and principled way to learn better decision rules for improving computer systems is an active area of research. Recently, many have proposed to use supervised learning and reinforcement learning as potential solutions, under the umbrella term \emph{machine learning for systems}. In the query optimisation community, all of these proposals function in a batch regime, whereby the model is static and has to be retrained to exploit newly available information. Meanwhile, online machine learning allows to learn from a stream of data, and is thus able to remain up-to-date. Indeed, many computer systems are event-based applications, which means that algorithms which are able to process event streams might have some edge over static algorithms. This is even more so important in the case where query workloads are modified or when the underlying data distribution changes.

As an example, in this paper we focus on the task of selectivity estimation in database cost models. Selectivity estimation is a difficult problem whereby the cost model has to predict how many tuples will be produced a particular query execution plan. There have recently been proposals that explore the use of deep learning to solve it. Said proposals follow the batch paradigm, and do not offer a principled answer to take into account concept drift, such as workload modifications. Instead, we advocate the use of online machine learning. Model that obey this paradigm learn on the fly and as such do not have to be retrained from scratch. An added benefit is that they allow handling concept drift. Moreover, online models are more lightweight than their batch counterparts, which is of importance in a resource intensive environment, such as a database system.

We back-up our proposal by experimenting on the JOB benchmark. We simulate a workload by randomly sampling query execution plans from PostgreSQL. We ask models to predict the selectivity of each execution plan and measure the $q$-error. The online models are updated every time a QEP terminates. Meanwhile, the batch models are warmed-up prior to the workload and are thus static. Our experiments show that the online models are competitive with batch models. Moreover, online models are able to adapt when the query workload changes, be it in a hard or a soft manner. On the whole, online machine learning is a promising approach to improving selectivity estimation, and remains somewhat unexplored. More generally, we believe that online machine learning can and should be used for other applications where heuristic rules are applied and feedback is constantly streaming in.

\bibliographystyle{unsrt}  


\appendix

\section{Drift-resilient Bayesian linear regression} \label{robust-bln}

The emphasis of this paper isn't so much on the choice of the online machine learning models as much as this is on the basic principle of using online machine learning. Therefore, we have resorted to using methods that are established in the statistical learning community. However, we also stumbled on a twist to Bayesian linear regression which provided us with good experimental results. Bayesian modeling is a framework for mixing prior knowledge with observed evidence. In the case of linear regression, we can impose a prior distribution on the weights. We can denote this prior distribution as $p(\theta)$. A typical parametrisation choice is to use a multivariate normal distribution centered in 0. In such a case, it may seem at first that the prior is \emph{uninformative} because it is vague and doesn't contain any subjective information. However, this prior becomes very useful in a streaming context.

A Bayesian model can be updated with observed samples. The goal is to adjust the parameters of the model according to the observed information, whilst taking into account the current knowledge. In some sense, this goal is shared with that of online machine learning. The advantage of Bayesian modeling is that it offers update formulas which are consistent with the rules of probability, and are in fact optimal under the latter. Given a new sample $(x_t, y_t)$, Bayesian modeling gives us a mechanism for obtaining a new parameter distribution $p(\theta_{t+1} \mid \theta_t, x_t, y_t)$, which is therefore conditioned on the current distribution and the new sample. This probability distribution is obtained via Bayes' rule, as so:

\begin{equation}
    p(\theta_{t+1} \mid \theta_t, x_t, y_t) = \frac{p(y_t \mid x_t, \theta_t) \times p(\theta_t)}{p(\theta_t, x_t, y_t)}
\end{equation}

The left-hand side of the numerator is the likelihood of observing $y_t$ given the current parameter distribution $\theta_t$ and the features $x_t$. The right-hand side is the current  parameter distribution. The denominator is the distribution, which isn't in fact known. However, because it isn't dependent on $\theta$, it can be simplified depending on the chosen parametrisation. In fact, the mathematical details work nicely when the prior distribution and the likelihood are said to be \emph{conjugate}. The latter is a mathematical term that describes the fact that two distributions can be fused into a new distribution with updated parameters. When this isn't the case, then one has to resort to approximate Bayesian inference, which is beyond the scope of this discussion. One way to see it is that we are interested in the ``old-school'' way of doing Bayesian modeling, whereby distributions are conjugate to each other, which leads to analytical formulas that are well suited to online machine learning. To keep things general, we will simply write down:

\begin{equation}
    p(\theta_{t+1} \mid \theta_t, x_t, y_t) \propto p(y_t \mid x_t, \theta_t) p(\theta_t)
\end{equation}

The previous statement simply expresses the fact that the posterior distribution of the model parameters is proportional to the product of the likelihood and the prior distribution. In other words, in can be obtained using an analytical formula that is specific to the chosen likelihood and prior distribution. If we're being pragmatic, then what we're really interested in is to obtain the \emph{predictive distribution}, which is obtained by marginalising over the model parameters $\theta_t$:

\begin{equation}
    p(y_t \mid x_t) = \int p(y_t \mid \textbf{w}, x_t) p(\textbf{w}) d\textbf{w}
\end{equation}

Again, this isn't analytically tractable, except if the likelihood and the prior are conjugate to each other. The equation does make sense though, because it expressed the fact we're computing a weighted average of the potential $y_i$ values for each possible model parameter $\textbf{w}$, therefore accounting for our uncertainty in the weight parameters.

\begin{equation}
    p(y_t \mid x_t) \propto p(y_t \mid x_t, \theta_t) p(\theta_t)
\end{equation}

In short, the predictive distribution can be obtained by mixing the predictive distribution and the current parameter distribution. Again, this isn't analytically tractable, except if the likelihood and the prior are conjugate to each other.

For the purpose of online machine learning, what matters is that we can update the distribution of the parameters when a new pair $(x_t, y_t)$ arrives:

\begin{equation}
    p(\theta_{t+1} \mid \theta_t, x_t, y_t) \propto p(x_t, y_t \mid \theta_t) p(\theta_t)
\end{equation}

Before any data comes in, the model parameters follow the initial distribution we picked, which is $p(\theta_0)$. At this point, if we're asked to predict $y_0$, then it's predictive distribution would be obtained as so:

\begin{equation}
p(y_0 \mid x_0) \propto p(y_0 \mid x_0, \theta_0) p(\theta_0)
\end{equation}

Next, once the first observation $(x_0, y_0)$ arrives, we can update the distribution of the parameters:

\begin{equation}
p(\theta_1 \mid \theta_0, x_0, y_0) \propto p(x_0, y_0 \mid \theta_0) p(\theta_0)
\end{equation}

The predictive distribution, given a set of features $x_1$, is thus:

\begin{equation}
p(y_1 \mid x_1) \propto p(y_1 \mid x_1, \theta_1) \underbrace{p(\theta_1 \mid \theta_0, x_0, y_0)}_{p(\theta_1)}
\end{equation}

The previous equations expresses the fact that the prior of the weights for the current iteration is the posterior of the weights at the previous iteration. Once the second pair $(x_1, y_1)$ is available, the distribution of the model parameters is updated in the same way as before:

\begin{equation}
p(\theta_2 \mid \theta_1, x_1, y_1) \propto p(y_1 \mid x_1, \theta_1) \underbrace{p(y_0 \mid x_0, \theta_0) p(\theta_0)}_{p(\theta_1)}
\end{equation}

When the pair $(x_2, y_2)$ arrives, the distribution of the weights can be obtained once again:

\begin{equation}
p(\theta_3 \mid \theta_2, x_2, y_2) \propto p(y_2 \mid x_2, \theta_2) \underbrace{p(y_1 \mid x_1, \theta_1) \underbrace{p(y_0 \mid x_0, \theta_0) p(\theta_0)}_{p(\theta_1)}}_{p(\theta_2)}
\end{equation}

By now, it might be clear that there is recursive relationship that links each iteration: the posterior distribution at step $t$ becomes the prior distribution at step $t+1$. This simple fact is the reason why analytical Bayesian inference can naturally be used as an online machine learning algorithm. Indeed, we only need to store the current distribution of the weights to make everything work.

Up until now we didn't give any useful example. We will now see how to perform linear regression by using Bayesian inference. In a linear regression, the model parameters $\theta_t$ are just weights $w_t$ that are linearly applied to a set of features $x_t$:

\begin{equation}
    y_t = w_t x_t^\intercal + \epsilon_t
\end{equation}

Each prediction is the scalar product between $p$ features $x_t$ and $p$ weights $w_t$. The trick here is that we're going to assume that the noise $\epsilon_i$ follows a given distribution. In particular, we will be boring and use the Gaussian ansatz, which implies that the likelihood function is a Gaussian distribution:

\begin{equation}
    p(y_t \mid x_t, w_t) = \mathcal{N}(w_t x_t^\intercal, \beta^{-1})
\end{equation}

Christopher Bishop calls $\beta$ the ``noise precision parameter''. In statistics, the precision is inversely related to the noise variance as so: $\beta = \frac{1}{\sigma^2}$. Basically, it translates our belief on how noisy the target distribution is. Both concepts coexist mostly because statisticians can't agree on a common Bible. There are ways to tune this parameter automatically from the data, however for the sake of simplicity we will treat it as known constant. In any case, the appropriate prior distribution for the above likelihood function is the multivariate Gaussian distribution:

\begin{equation}
    p(w_0) = \mathcal{N}(m_0, S_0)
\end{equation}

$m_0$ is the mean vector of the distribution while $S_0$ is its covariance matrix. Initially, their initial values will be:

\begin{equation}
    m_0 = (0, \dots , 0)
\end{equation}

\begin{equation}
    S_0 = \begin{pmatrix} \alpha^{-1} & \dots & \dots \\\ \dots & \alpha^{-1} & \dots \\\ \dots & \dots & \alpha^{-1} \end{pmatrix}
\end{equation}

The value $\alpha$ is a hyperparameter that needs to be provided. From our experience, its influence is very small in an online scenario and therefore its value does not matter very much. We can now determine the posterior distribution of the weights:

\begin{equation}
p(w_{t+1} \mid w_t, x_t, y_t) = \mathcal{N}(m_{t+1}, S_{t+1})
\end{equation}

\begin{equation}
S_{t+1} = (S_t^{-1} + \beta x_t^\intercal x_t)^{-1}
\end{equation}

\begin{equation}
m_{t+1} = S_{t+1}(S_t^{-1} m_t + \beta x_t y_t)
\end{equation}

Note that $x_t^\intercal x_t$ is the outer product of $x_t$ with itself. There are also a set of formulas that can be used to obtain the predictive distribution:

\begin{equation}
p(y_t) = \mathcal{N}(\mu_t, \sigma_t)
\end{equation}

\begin{equation}
\mu_t = w_t x_t^\intercal
\end{equation}

\begin{equation}
\sigma_t = \frac{1}{\beta} + x_t S_t x_t^\intercal
\end{equation}

All of the above formulas are quite common and can be found in many introductions to Bayesian inference. One of the issues with this formulation is that the data is assumed to be stationary. Indeed, the more data we show the model, the more it will be confident about its parameter estimate. However, we could like to it to be able to be robust to concept drift by forgetting the past and focusing on recent data. The solution we found was to change the update formulas of the covariance matrix and the mean vector in the following manner:

\begin{equation}
S_{t+1} = (\gamma S_t^{-1} + (1 - \gamma)\beta x_t^\intercal x_t)^{-1}
\end{equation}

\begin{equation}
m_{t+1} = S_{t+1}(\gamma S_t^{-1} m_t + (1 - \gamma) \beta x_t y_t)
\end{equation}

In the above equations, $\gamma$ acts a smoothing parameter which controls how much the model ``forgets'' its current state and sticks to the new data. In the case where $\gamma = 1$, the model doesn't learn and sticks to the prior distribution. On the contrary, when $\gamma = 0$, the model memorises the latest sample and forgets what it has seen up to there. These formulas are very much heuristic and we have not taken the time to give them a thorough analytical treatment. As far as we can tell, they haven't been used in published literature. However, they are not complex and resemble exponential weighted moving averages. In practice, we have found that this formulation provides a robust off-the-shelf algorithm that works well on average for many problems where concept drift occurs. In our experience, the importance of choosing a suitable $\gamma$ doesn't matter very much. Indeed, we advise using $\gamma = 0.7$ by default.

\end{document}